\documentclass[runningheads]{llncs}
\usepackage{amsmath}
\usepackage{fontawesome}
\usepackage{textcomp}
\usepackage{bbm}

\usepackage[T1]{fontenc}
\usepackage{graphicx}
\begin{document}
\title{Improving Automatic Fetal Biometry Measurement with Swoosh Activation Function\thanks{This research was supported by Australian Research Council (ARC) DP200103748.}}
\author{Shijia Zhou\inst{1,2}\textsuperscript{(\faEnvelopeO)}\orcidID{0000-0002-5936-8948}  \and
Euijoon Ahn\inst{3}\orcidID{0000-0001-7027-067X} \and
Hao Wang\inst{1}\orcidID{0000-0002-6235-8425} \and
Ann Quinton\inst{4}\orcidID{0000-0001-6585-7468} \and
Narelle Kennedy\inst{5}\orcidID{0000-0003-0801-6317} \and
Pradeeba Sridar\inst{6}\orcidID{0000-0003-1191-5188} \and
Ralph Nanan\inst{7}\orcidID{0000-0002-1749-6378} \and
Jinman Kim\inst{1}\orcidID{0000-0001-5960-1060}
} 
%index{Zhou, Shijia}
%index{Ahn, Euijoon}
%index{Wang, Hao}
%index{Quinton, Ann}
%index{Kennedy, Narelle}
%5index{Sridar, Pradeeba}
%index{Nanan, Ralph}
%index{Kim, Jinman}
\authorrunning{S. Zhou et al.}

\institute{University of Sydney, School of Computer Science, Sydney NSW 2000, Australia \and
James Cook University, Smithfield QLD 4878, Australia \and
Central Queensland University, Sydney NSW 2000, Australia \and
Liverpool Hospital, Liverpool NSW 2170, Australia \and
Indian Institute of Technology, Delhi, New Delhi, Delhi 110016, India \and
University of Sydney, Sydney Medical School Nepean, Kingswood NSW 2747, Australia
}
\titlerunning{Improving Automatic Fetal Biometry Measurement}
\maketitle              % typeset the header of the contribution
\begin{abstract}
The measurement of fetal thalamus diameter (FTD) and fetal head circumference (FHC) are crucial in identifying abnormal fetal thalamus development as it may lead to certain neuropsychiatric disorders in later life. However, manual measurements from 2D-US images are laborious, prone to high inter-observer variability, and complicated by the high signal-to-noise ratio nature of the images. Deep learning-based landmark detection approaches have shown promise in measuring biometrics from US images, but the current state-of-the-art (SOTA) algorithm, BiometryNet, is inadequate for FTD and FHC measurement due to its inability to account for the fuzzy edges of these structures and the complex shape of the FTD structure. To address these inadequacies, we propose a novel Swoosh Activation Function (SAF) designed to enhance the regularization of heatmaps produced by landmark detection algorithms. Our SAF serves as a regularization term to enforce an optimum mean squared error (MSE) level between predicted heatmaps, reducing the dispersiveness of hotspots in predicted heatmaps. Our experimental results demonstrate that SAF significantly improves the measurement performances of FTD and FHC with higher intraclass correlation coefficient scores in FTD and lower mean difference scores in FHC measurement than those of the current SOTA algorithm BiometryNet. Moreover, our proposed SAF is highly generalizable and architecture-agnostic. The SAF's coefficients can be configured for different tasks, making it highly customizable. Our study demonstrates that the SAF activation function is a novel method that can improve measurement accuracy in fetal biometry landmark detection. This improvement has the potential to contribute to better fetal monitoring and improved neonatal outcomes.

\end{abstract}
%
%
%
%
%
%
%Certain maternal microbiome bacteria have also been found to aid in normal fetal neurodevelopment \cite{Meckel2020MaternalWiring}
\section{Introduction}
The thalamus is a critical brain region that relays and modulates information between different parts of the cerebral cortex, and plays a vital role in signal transmission and processing, including pain recognition and reaction \cite{Waxman2020ChapterHypothalamus}. Abnormal fetal thalamus development, which can disrupt serotonin receptor development, may contribute to the development of later neuropsychiatric disorders \cite{Wai2011ProfilesThalamus}. Thalamus development is influenced by maternal factors such as diabetes that suppresses thalamus development after 2 weeks of gestation \cite{You2022GLP-1Development}. To further investigate the relationship between maternal factors and fetal thalamus growth, a dataset containing maternal factors, fetal thalamus diameter (FTD), and fetal head circumference measurements (FHC) is needed. FHC measurement is necessary to normalize FTD against gestational age. However, measuring FTD and FHC manually from 2D-US scans is laborious, prone to high inter-observer variability, and complicated by 2D-US images' high signal-to-noise ratio (SNR) nature due to ultrasound wave's lack of penetration power compared to ionizing radiations such as the x-ray, acoustic shadows cast by highly echogenic objects, and unique noise characteristics due to ultrasound reverberation \cite{Bethune2013AUltrasound,Sridar2017AutomaticSets,Brickson2021ReverberationNetwork}.

In recent years, landmark-based detection approaches based on deep learning have been employed to measure biometrics from fetal US images. They were used to detect measurement key points for brain structures in the fetal brains, and bony structures such as length of the femur or dimensions in the pelvic floors \cite{avisdris2022biometrynet,Shankar2022LeveragingBrain,Xia2022AutomaticUltrasound}. In these studies, the distance between a pair of landmarks represents the biometry being measured. The current state-of-the-art (SOTA) landmark detection algorithm for measuring biometry in fetal 2D-US images is the BiometryNet proposed by Avisdris et al., which was developed to detect landmarks for measuring the FHC and fetal femur length \cite{avisdris2022biometrynet}. BiometryNet is based on High-Resolution Net (HRNet) and it has shown great performances in measuring dimensions of fetal skull and femur bone, outperforming other landmark-based methods \cite{sun2019deep}. BiometryNet uses dynamic orientation determination (DOD) to enforce a consistent orientation between detected landmarks', i.e. the first landmark is always the left/top measurement key point, and the second landmark is always the right/bottom key point \cite{avisdris2022biometrynet}.

However, BiometryNet cannot be directly used to measure FTD and FHC due to two specific difficulties. First being that the "guitar-shaped" structure (GsS) by Sridar et al. to measure FTD has similar echogenicity to surrounding brain tissues, resulting in fuzzy boundaries, especially around the wing-tips where measurement landmarks are located \cite{Sridar2017AutomaticSets} (Fig.~\ref{fig:head, femur, thalamus}). This difficulty is also observed in 2D-US images of fetal skulls when they appear broken due to unfused bones and acoustic shadows cast by the skull bones themselves (Fig.~\ref{fig:head, femur, thalamus}). Second being that the shape of the GsS resembles the silhouette of a guitar and more complex than that of a skull or femur bone (Fig.~\ref{fig: GsS_landmarks_heatmaps}) \cite{Sridar2017AutomaticSets}. These two difficulties causes uncertainties in the localization of measurement landmarks of FTD, resulting in inaccurate measurement of FTD.

% The direct application of BiometryNet for measuring FTD faces two challenges. Firstly, DOD does not account for the high SNR nature of 2D-US images. This is particularly problematic when measuring FTD since the proposed "guitar-shaped" structure (GsS) by Sridar et al. has similar echogenicity to surrounding brain tissues, resulting in fuzzy boundaries, especially around the wing-tips where measurement landmarks are located \cite{Sridar2017AutomaticSets} (Fig.~\ref{fig:head, femur, thalamus}). This can be observed in either early pregnancy when the skull is not fully mineralized, or mid-late pregnancy when the skull appears broken due to unfused bones and acoustic shadows cast by the skull bones themselves (Fig.~\ref{fig:head, femur, thalamus}). Conversely, the fetal femur is a bone structure with distinct echogenicity that differs from surrounding muscle tissues and does not appear broken in fetal 2D-US images (Fig.~\ref{fig:head, femur, thalamus}). 

{
\begin{figure}
    \centering
    \includegraphics[width=\textwidth]{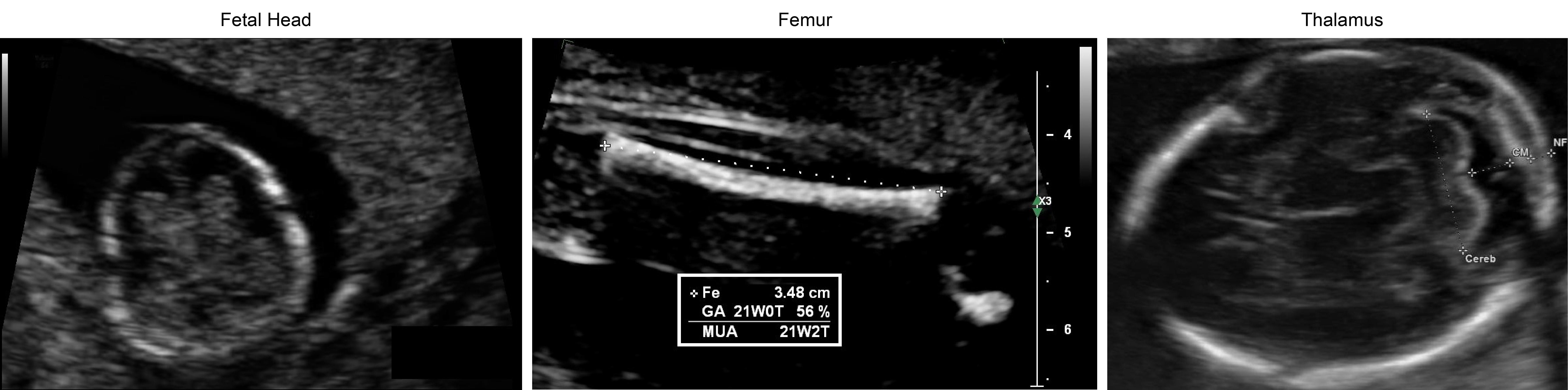}
    \caption{Left: 2D-US image of fetal head from the HC18 dataset \cite{ThomasL.A.vandenHeuvel2018AutomatedImages}, note part of the skull is not mineralized and has similar echogenicty to the adjacent uterine tissue. Middle: 2D-US image of a fetal femur \cite{ApostolosKolitsidakis2021HowLength}. Right: 2D-US image of a GsS for measuring FTD, note the gaps in the skull due to unfused bones.}
    \label{fig:head, femur, thalamus}
\end{figure}
}

%When the input to the function is greater than 0, the SAF minimizes to a pre-defined point. On the left side of the minimum point, SAF grows exponentially while being constrained by the y-axis as an asymptote. On the right side of the minimum point, SAF is constrained by a power function with a predetermined exponent, resulting in a gentler growth rate.
To address the above difficulties, we present a novel Swoosh Activation Function (SAF) designed to enhance the regularization of heatmaps produced by landmark detection algorithms. The SAF takes its name from the Nike\texttrademark\space swoosh logo, which resembles its shape (Supplementary Fig. 1). SAF can serve as a regularization term to enforce an optimum mean squared error (MSE) level between a pair of predicted heatmaps ($pHs$). By doing so, the landmark detection algorithm is compelled to highlight different areas. Additionally, SAF can enforce an optimum MSE between individual $pH$ and a zero matrix to reduce hotspot dispersiveness. Moreover, because SAF does not grow exponentially when the input MSE is higher than the optimum MSE level, it does not hinder algorithm's learning. Consequently, we hypothesize that SAF can enhance landmark detection accuracy and overcome uncertainties arising from the fuzzy edges of GsS by promoting hotspot concentration in $pHs$. 

\section{Method}
%SAFNet is a landmark detection algorithm based on BiometryNet. Compared to BiometryNet, it utilizes SAF in its loss function to improve heatmap prediction for measuring FTD and FHC. For each measurement, SAFNet outputs two heatmaps, one for each landmark. The hottest point on the heatmap corresponds to the predicted landmark (Fig.~\ref{fig: GsS_landmarks_heatmaps}.C,D). We also implemented SAF in a different landmark detection network based on EfficientNet in its b4 configuration and the transpose convolution-based heatmap prediction algorithm proposed by Xiao et al. \cite{Tan2019EfficientNet:Networks,Xiao2018SimpleTrackingb}. We simply name this comparison network EfficientNet.

\subsection{Swoosh Activation Function}
SAF is introduced to optimize $pHs$ by enforcing an optimum MSE between a pair of $pHs$ and a secondary optimum MSE between a predicted heatmap ($pH$) and a zero matrix ($O$). We determine the optimum MSE by computing the MSE between a pair of ground truth heatmaps ($gHs$). Each ground truth heatmap ($gH$) represents a measurement landmark by a smaller matrix drawn from a Gaussian distribution that is centered at the landmark coordinates with the peak assigned at the value of 1 (Fig.~\ref{fig: MSEs of good and  bad heatmap pairs}.A). Since we determine that $gHs$ represent optimum heatmaps, the MSE between a pair of $pHs$ should approximate the MSE between a pair of $gHs$. In addition, the secondary optimum MSE between a $pH$ and a zero matrix is half of the MSE between a pair of $gHs$, since only one Gaussian distribution is being compared. We demonstrate in Fig.~\ref{fig: MSEs of good and bad heatmap pairs}.B and C how deviations from this optimum MSE value can lead to incorrect and noisy heatmaps.
{
\begin{figure}
    \centering
    \includegraphics[width=\textwidth]{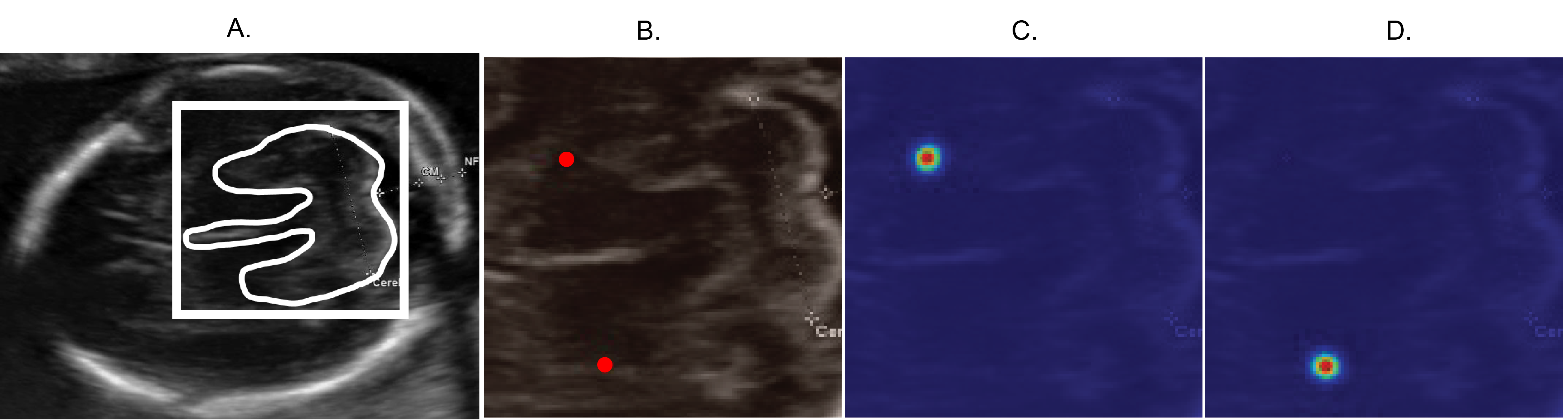}
    \caption{A: a 2D-US fetal brain image that has the GsS (outlined with white curly lines) annotated with manually created bounding box (white box); B: input image of the GsS constrained by the bounding box. The red dots represent the ground truth landmarks of FTD; C: the first heatmap of one of the FTD landmarks with the hottest (red) spot representing the landmark; D: the second heatmap of the other FTD landmark with the hottest (red) spot representing the landmark}
    \label{fig: GsS_landmarks_heatmaps}
\end{figure}
}

To enforce this pre-determined optimum MSE, we defined SAF as:
\begin{equation}
f(x>0) = \left(ax + \frac{1}{bx}\right)^c - Min
\label{equation for SAF}
\end{equation}
In Eq. \ref{equation for SAF}, $Min$ represents a function that ensures SAF minimizes to 0, and it is defined as $Min = f\left(\sqrt{\frac{1}{ab}}\right)$. The coefficient $a$ determines the slope of SAF around the minimum point in Quadrant 1 of the Cartesian coordinate system (Supplementary Fig. 1). The slope of SAF determines its regularization strength. Coefficient $b$ determines the $x$-axis coordinate of the minimum point where the x-coordinate of the minimum point correspond to the optimum MSE. Coefficient $b$ is be deducted by Supplementary Eq. 1 and the coefficient $c$ is deducted by Supplementary Eq. 2.

\subsection{SAF Regularization}
The loss function consists of the MSE between $pH$ and $gH$, and three additional SAF regularization terms. We chose SAF as the activation function to control these regularization terms because SAF's output grows exponentially on either side of the minimum point (Supplementary Fig. 1) where the x-axis represents input values of $MSE(pH_1,pH_2)$, $MSE(pH_1, O)$, and $MSE(pH_2, O)$, while y-axis represents the output values of SAF which minimize to 0 when input values approximate the predetermined optimum MSE. SAF locks the input MSE to the pre-determined optimum value because deviation from this value would result in a fast increase in the gradient of SAF. The first SAF term regularizes the MSE between a pair of $pHs$. The next two SAF terms regularize the MSE between each $pH$ and a zero matrix. The equation for the entire loss function is:
    \begin{equation}
    \begin{split}
    L(pH_1,pH_2,gH_1,gH_2, O) & = MSE(pH_1,gH_1) + MSE(pH_2,gH_2) + \\ 
    & SAF(MSE(pH_1,pH_2)) + SAF(MSE(pH_1, O)) + \\ 
    & SAF(MSE(pH_2, O))
    \end{split}
    \label{Hetmap Predict Loss Function}
    \end{equation}
where $pH_1 = $ the first predicted heatmap; $pH_2 = $ the second predicted heatmap; $gH_1 = $ the first ground truth heatmap; $gH_2 = $ the second ground truth heatmap; $O = $ a zero matrix where the elements are zero.

{\begin{figure}
    \centering
    \includegraphics[width=0.7\textwidth]{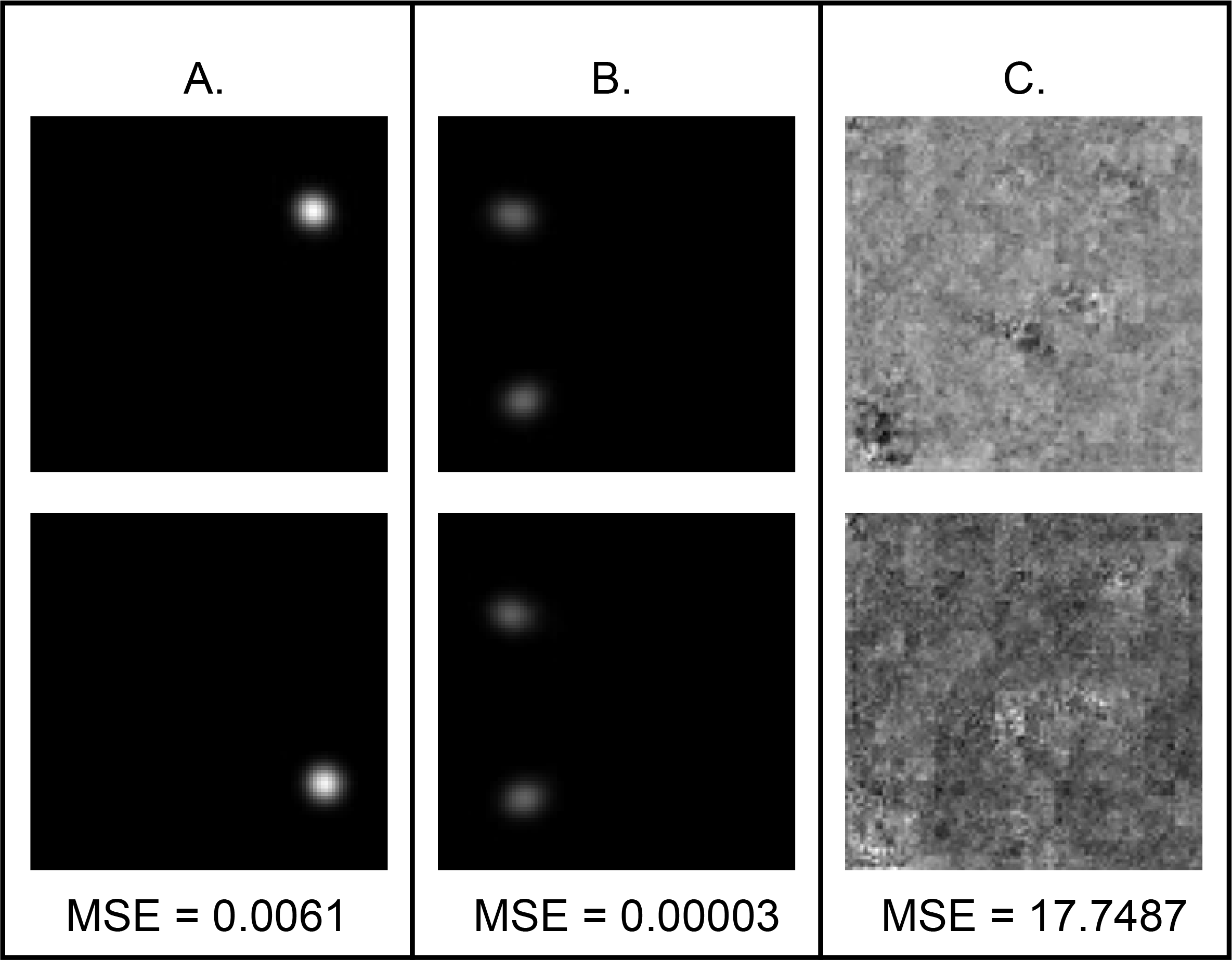}
    \caption{A (left column): a pair of ground truth heatmaps with optimum MSE = 0.0061. B (middle column): a pair of predicted heatmaps with low MSE = 0.00003. C (right column): a pair of predicted heatmaps with high MSE = 17.7487}
    \label{fig: MSEs of good and  bad heatmap pairs}
\end{figure}}

\subsection{Datasets}
\subsubsection{FTD Dataset:} The dataset used in this study consists of 1111 2D-US images acquired during the second trimester of pregnancies and confirmed by board-certified ultrasonographers to be suitable for measuring FTD \cite{Sridar2020NormativeExaminations}. No additional ethics approval was required. Spatial constraints were provided by manually added bounding boxes around the GsS, verified by the same ultrasonographers. Pycocotools generated two $gHs$ for each pair of measurement landmarks, with hotspots representing landmarks \cite{lin2014microsoft}. 5-fold cross-validation was performed. During training, 100 training samples were randomly held out as the validation set. The intraclass correlation coefficient (ICC) score was computed using IBM\texttrademark\space SPSS\texttrademark\space version 28, with the ICC configuration being Two-Way Random and Absolute Agreement\cite{IBMCorp2020IBMWindows}.
\subsubsection{HC18 Dataset:} HC18 dataset is available on the Grand Challenge website \cite{ThomasL.A.vandenHeuvel2018AutomatedImages}. We utilized least squared fitting of an ellipse to determine the center, width, height, and angle of rotation of the elliptical ground truth mask. We used trigonometry to determine the coordinates of the landmarks for the major and minor axes of the ellipse. Spatial constraints in the form of bounding boxes were built by the ground truth mask of head circumference published by the dataset author. Predicted landmark points were used for the major and minor axes to calculate the ellipse width and height for testing. The predicted major and minor axes were assumed to be perpendicular, and their point of intersection was used as the center of the ellipse. Finally, we uploaded the results to the Grand Challenge leaderboard and used the mean difference between predicted FHC and ground truth as the performance metric.

\subsection{Experimental Setup}

\subsubsection{Training Epochs and Learning Rate}

We conducted our experiments using PyTorch version 1.12 on two NVIDIA GTX-1080Ti graphical processing units, each with 11GB of video memory. For landmark detection training, we used the same learning rate configuration as Avisdris et al. \cite{avisdris2022biometrynet} to train both BiometryNet with / without SAF. Specifically, we set the initial learning rate to 10\textsuperscript{-5} and reduced it using a multi-step learning rate scheduler that scaled the learning rate by a factor of 0.2 at epoch numbers 10, 40, 90, and 150. We trained all BiometryNet models for a total of 200 epochs.
For EfficientNet with / without SAF, we set the initial learning rate to 10\textsuperscript{-5} and reduced it using a multi-step learning rate scheduler that scaled the learning rate by a factor of 0.1 at epoch number 300 and 350. We trained both models for 400 epochs.
\subsubsection{Pre-processing}

Our pre-processing pipeline included random rotation of ± 180 degrees, random re-scaling of ± 5\%, resizing to $384 \times 384$ pixels without preserving aspect ratio, and normalization using ImageNet-derived mean = (0.485, 0.456, 0.406) and standard deviation = (0.229, 0.224, 0.225) for each color channel.

\subsubsection{SAF Configuration}
Given our dataset configuration where each biometry landmark was represented by a $19 \times 19$ matrix with values derived from a Gaussian distribution with the center point's value peaked at 1, the MSE between a pair of $gHs$ was 0.0061. This configuration followed the standard implementation used in human pose estimation landmark detection \cite{xiao2018simple}. The secondary optimum MSE between a $pH$ and a zero matrix was halved at 0.00305. We also predetermined the value of $Min$ to be 0.001 to prevent SAF from overpowering the MSE loss between $gHs$ and $pHs$. We experimented with different values of the coefficient $a$ (1, 4, and 8) to evaluate coefficient $a$'s influence on landmark detection accuracy. We then determined the value of coefficient $b$ using Supplementary Eq. 1, and the value of coefficient $c$ using Supplementary Eq. 2. 

We conditionally activated SAF when the average of $MSE(pH_1,gH_1)$ and $MSE(pH_2,gH_2)$ was less than 0.0009 because SAF is not bounded and early activation would hinder algorithm learning. The proposed SAF algorithm was evaluated using six model configurations, including Vanilla BiometryNet, BiometryNet with SAF with coefficient $a$ values of 1, 4, and 8, an EfficientNet, and the EfficientNet with SAF configured with coefficient $a$ value of 4. The model configurations were trained and tested to verify the usefulness of the proposed SAF using both FTD and HC18 datasets.
\section{Results}

\begin{figure}
    \centering
    \includegraphics[width=0.7\textwidth]{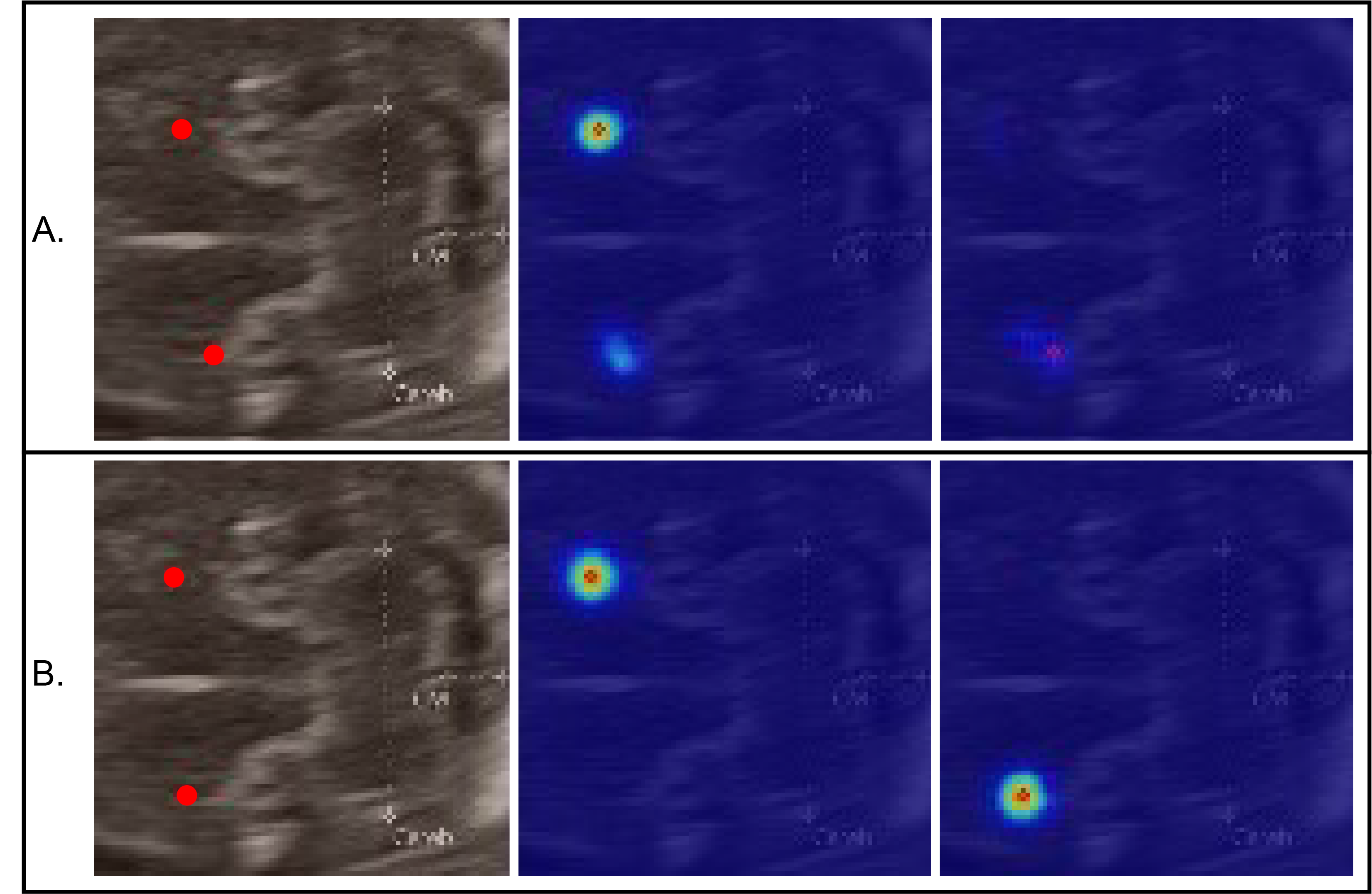}
    \caption{Row A: landmarks predicted and heatmaps produced by BiometryNet. Left, input image overlaid with predicted landmarks (red spots). Middle, the first predicted heatmap, there are hotspots present near both upper and lower wing-tips. Right, the second predicted heatmap. Row B: landmarks predicted and heatmaps produced by BiometryNet\_SAF\_a1. Left input image overlaid with predicted landmarks (red spots). Middle, the first predicted heatmap. Right, the second predicted heatmap.}
    \label{fig:heatmap comparison}
\end{figure}

The results of FTD dataset show that BiometryNet with SAF\_a1 (BiometryNet\_SAF\_a1) achieved the highest ICC score at 0.737, surpassing the performance of the vanilla BiometryNet, which scored 0.684. Moreover, BiometryNet with SAF\_a4 (BiometryNet\_SAF\_a4) and BiometryNet with SAF\_a8 (BiometryNet\_SAF\_a8) also demonstrated superior ICC scores for FTD measurement, albeit to a lesser degree than BiometryNet\_SAF\_a1. The impact of SAF on performance was further observed in the modified EfficientNet, where EfficientNet\_SAF\_a4 achieved a higher ICC score of 0.725, compared to the modified EfficientNet without SAF, which only scored 0.688. We also observed that SAF reduced the similarities between a pair of $pHs$ and the dispersiveness of hotspot in the $pH$. We display such heatmaps produced by BiometryNet in Fig~.\ref{fig:heatmap comparison}.A and BiometryNet\_SAF\_a1 in Fig~.\ref{fig:heatmap comparison}.B.

For the HC18 dataset, BiometryNet\_SAF\_a8 demonstrated the lowest measurement mean difference from the ground truth at 3.86 mm ± 7.74 mm. Additionally, all configurations outperformed the vanilla BiometryNet. The impact on FHC measurement was further observed in the modified EfficientNet, where EfficientNet\_SAF\_a4 achieved a lower measurement mean difference at 4.87 mm ± 5.79 mm compared to EfficientNet at 32.76 mm ± 21.01 mm. The FTD dataset ICC scores and HC18 dataset mean differences for each algorithm are presented in Table~\ref{results table}.
{\begin{table}
\centering
\caption{ICC scores of the FTD dataset, and mean measurement differences ± confidence interval (CI) of the HC18 dataset achieved by all the algorithms.}\label{results table}
\begin{tabular}{|l|l|l|}
\hline
Network Name & FTD ICC {\hskip 0.1in} & HC18 Mean Difference ± CI (mm) {\hskip 0.1in}\\
\hline
Vanilla BiometryNet &  0.684 & 4.56 ± 7.41\\
BiometryNet\_SAF\_a1 & 0.737 & 4.02 ± 6.70\\
BiometryNet\_SAF\_a4 &  0.724 & 4.03 ± 7.73\\
BiometryNet\_SAF\_a8 & 0.719 & 3.86 ± 7.74\\
EfficientNet & 0.688 & 32.76 ± 21.01\\
EfficientNet\_SAF\_a4 & 0.725 & 4.87 ± 5.79\\
\hline
\end{tabular}
\end{table}}
\section{Discussion}
The main findings of this study are as follows: (1) SAF improved the measurement accuracy of algorithms in both FTD and FHC measurement tasks; (2) SAF regularization is architecture-agnostic, as it improved the measurement accuracy of both BiometryNet and EfficientNet compared to their vanilla forms that do not use SAF; and (3) the optimum configuration of SAF coefficients is task-dependent. For FTD measurement, the most optimum configuration was to use $a = 1$, while for FHC measurement was with coefficient $a = 8$.

The results demonstrate the performance improvement brought about by SAF regularization, effectively improving the accuracy of landmark detection of fetal biometries in 2D-US images. SAF regularization forces a pair of heatmaps to highlight different areas and reduce the dispersiveness of hotspots in $pHs$, which results in improved fetal biometry landmark detection accuracy. This is supported by the comparison of heatmaps produced by BiometryNet and BiometryNet\_SAF\_a1 displayed in Fig.~\ref{fig:heatmap comparison}. Moreover, SAF regularization is simple to implement and easy to configure, requiring no modification to the network architecture. Our results also suggest that SAF is highly generalizable as it is architecture-agnostic, improving the performance of both BiometryNet and EfficientNet. SAF is also highly configurable for different tasks via different coefficient configurations. Furthermore, we also suggest that SAF is generalizable to other imaging modalities that also require pair-wise landmark detection. For example, detecting mitral and aortic valves in 2D-US images of heart and detecting cranial sutures in CT images of skull.

As part of our future study, we will explore the effectiveness of SAF regularization in other fetal landmark detection tasks, especially those that suffer from similar issues with fuzzy edges and uncertain landmark locations. Additionally, the optimum configuration of SAF coefficients may vary depending on the specific dataset or imaging modality used.

\section{Conclusion}
Our study demonstrated the effectiveness of SAF as a novel activation function for regularizing heatmaps generated by fetal biometry landmark detection algorithms, resulting in improved measurement accuracy. SAF outperformed the previous state-of-the-art algorithm, BiometryNet, in both FTD and FHC measurement tasks. Importantly, our results showed that SAF is architecture-agnostic and highly configurable for different tasks through its coefficients, making it a generalizable solution for a wide range of landmark detection problems.


\bibliographystyle{splncs04}
\bibliography{Paper747}

\begin{thebibliography}{10}
\providecommand{\url}[1]{\texttt{#1}}
\providecommand{\urlprefix}{URL }
\providecommand{\doi}[1]{https://doi.org/#1}

\bibitem{ApostolosKolitsidakis2021HowLength}
{Apostolos Kolitsidakis}: {How to measure the femur length} (2021)

\bibitem{avisdris2022biometrynet}
Avisdris, N., Joskowicz, L., Dromey, B., David, A.L., Peebles, D.M., Stoyanov,
  D., Ben~Bashat, D., Bano, S.: Biometrynet: Landmark-based fetal biometry
  estimation from standard ultrasound planes. In: International Conference on
  Medical Image Computing and Computer-Assisted Intervention. pp. 279--289.
  Springer (2022)

\bibitem{Bethune2013AUltrasound}
Bethune, M., Alibrahim, E., Davies, B., Yong, E.: {A pictorial guide for the
  second trimester ultrasound}. Australasian journal of ultrasound in medicine
  \textbf{16}(3),  98--113 (2013). \doi{10.1002/j.2205-0140.2013.tb00106.x}

\bibitem{Brickson2021ReverberationNetwork}
Brickson, L.L., Hyun, D., Jakovljevic, M., Dahl, J.J.: {Reverberation Noise
  Suppression in Ultrasound Channel Signals Using a 3D Fully Convolutional
  Neural Network}. IEEE Transactions on Medical Imaging  \textbf{40}(4),
  1184--1195 (2021). \doi{10.1109/TMI.2021.3049307}

\bibitem{IBMCorp2020IBMWindows}
{IBM Corp}: {IBM SPSS Statistics for Windows} (2020)

\bibitem{lin2014microsoft}
Lin, T.Y., Maire, M., Belongie, S., Hays, J., Perona, P., Ramanan, D.,
  Doll{\'a}r, P., Zitnick, C.L.: Microsoft coco: Common objects in context. In:
  Computer Vision--ECCV 2014: 13th European Conference, Zurich, Switzerland,
  September 6-12, 2014, Proceedings, Part V 13. pp. 740--755. Springer (2014)

\bibitem{Shankar2022LeveragingBrain}
Shankar, H., Narayan, A., Jain, S., Singh, D., Vyas, P., Hegde, N., Kar, P.,
  Lad, A., Thang, J., Atada, J., Nguyen, D., Roopa, P.S., Vasudeva, A.,
  Radhakrishnan, P., Devalla, S.: {Leveraging Clinically Relevant Biometric
  Constraints to Supervise a Deep Learning Model for the Accurate Caliper
  Placement to Obtain Sonographic Measurements of the Fetal Brain}. In:
  Proceedings - International Symposium on Biomedical Imaging. vol. 2022-March.
  IEEE Computer Society (2022). \doi{10.1109/ISBI52829.2022.9761493}

\bibitem{Sridar2020NormativeExaminations}
Sridar, P., Kennedy, N.J., Quinton, A.E., Robledo, K., Kim, J., Nanan, R.:
  {Normative ultrasound data of the fetal transverse thalamic diameter derived
  from 18 to 22 weeks of gestation in routine second‐trimester morphology
  examinations}. Australasian journal of ultrasound in medicine
  \textbf{23}(1),  59--65 (2020). \doi{10.1002/ajum.12196}

\bibitem{Sridar2017AutomaticSets}
Sridar, P., Kumar, A., Li, C., Woo, J., Quinton, A., Benzie, R., Peek, M.J.,
  Feng, D., Kumar, R.K., Nanan, R., Kim, J.: {Automatic Measurement of Thalamic
  Diameter in 2-D Fetal Ultrasound Brain Images Using Shape Prior Constrained
  Regularized Level Sets}. IEEE journal of biomedical and health informatics
  \textbf{21}(4),  1069--1078 (2017). \doi{10.1109/JBHI.2016.2582175}

\bibitem{sun2019deep}
Sun, K., Xiao, B., Liu, D., Wang, J.: Deep high-resolution representation
  learning for human pose estimation. In: Proceedings of the IEEE/CVF
  conference on computer vision and pattern recognition. pp. 5693--5703 (2019)

\bibitem{ThomasL.A.vandenHeuvel2018AutomatedImages}
{Thomas L. A. van den Heuvel}, {Dagmar de Bruijn}, {Chris L. de Korte}, {Bram
  van Ginneken}: {Automated measurement of fetal head circumference using 2D
  ultrasound images} (2018)

\bibitem{Wai2011ProfilesThalamus}
Wai, M.S., Lorke, D.E., Kwong, W.H., Zhang, L., Yew, D.T.: {Profiles of
  serotonin receptors in the developing human thalamus}. Psychiatry Research
  \textbf{185}(1-2),  238--242 (1 2011). \doi{10.1016/j.psychres.2010.05.003}

\bibitem{Waxman2020ChapterHypothalamus}
Waxman, S.G., Waxman, S.G.: {Chapter 9: Diencephalon: Thalamus and
  Hypothalamus}. In: Clinical Neuroanatomy, chap.~9. McGraw-Hill Education, New
  York, 29th editi edn. (2020)

\bibitem{Xia2022AutomaticUltrasound}
Xia, W., Ameri, G., Fakim, D., Akhuanzada, H., Raza, M.Z., Shobeiri, S.A.,
  Mclean, L., Chen, E.C.: {Automatic Plane of Minimal Hiatal Dimensions
  Extraction From 3D Female Pelvic Floor Ultrasound}. IEEE Transactions on
  Medical Imaging  \textbf{41}(12),  3873--3883 (12 2022).
  \doi{10.1109/TMI.2022.3199968}

\bibitem{xiao2018simple}
Xiao, B., Wu, H., Wei, Y.: Simple baselines for human pose estimation and
  tracking. In: Proceedings of the European conference on computer vision
  (ECCV). pp. 466--481 (2018)

\bibitem{You2022GLP-1Development}
You, L., Deng, Y., Li, D., Lin, Y., Wang, Y.: {GLP-1 rescued gestational
  diabetes mellitus-induced suppression of fetal thalamus development}. Journal
  of Biochemical and Molecular Toxicology  (2022). \doi{10.1002/jbt.23258}

\end{thebibliography}

\end{document}